\documentclass  {article}
\usepackage[english, russian]{babel}
\begin{document}

\bigskip
\noindent {\Large \bf {Charge-energy with heat returns  back after the radial fall to gravitational center }}

\bigskip
\noindent{I.E. Bulyzhenkov, bulyzhenkov.ie@mipt.ru} 

\bigskip
\noindent{\it {Lebedev Physics Institute RAS\\
Moscow Institute of Physics and Technology, Moscow, Russia} }

\bigskip

{\small Abstract.
Heat transfer in the SR flow of power, rather than the Newton current of cold masses, provides proper referents for GR geodesic motion of inertial energy and for metric of non-empty space. GR can compare losses of internal energy under speed increases and can explain the attraction law by the body tendency toward equipartition of kinetic energies over internal and external degrees of freedom. Thermodynamic approach numerically describes the cyclic dynamics of the vertical fall to center with the final path deceleration followed by the accelerated rise as an oscillation around the kinetic energy equilibrium of geodetically moving body and pushes other testable predictions for inertial charges with heat in strong field gravitation.}

\bigskip




\section {Введение}
Во всех странах школьников приучают к догме, что релятивистская физика Эйнштейна в слабых полях должна воспроизводить физику Ньютона  с легко проверяемой динамикой постоянных масс для малых скоростей, $v^2\ll c^2$. При этом школьные учителя отмечают, что  классическая механика  точечных масс не знает численных неудач \lq\lq{}в области своей применимости\rq\rq{.}  Тем не менее, даже для самых малых скоростей механика Ньютона не может считаться правильной или полной теорией.  Ведь  классические массы не несут тепла (без которого на практике автомобиль не остановишь) и ньютоновскую динамику холодных масс в принципе необходимо дополнять термодинамикой и внешним термостатом даже в простейшей задаче двух тел.

В эйнштейновской физике инерционным и гравитационным зарядом является уже не масса, а полная энергия механического тела $E_o {\sqrt {g_{oo}}}/  {\sqrt {1 - v^2c^{-2}}}$ \cite {LL}, которая обязана учитывать тепло и его перенос по релятивистскому закону Лауэ 1908 года: $Q = Q_o {\sqrt {1 - v^2c^{-2}}}$. Именно так переносит свою энергию, т.е. частоту, и квантовая частица в волновой механике $\hbar \omega = \hbar \omega_o {\sqrt {1 - v^2c^{-2}}}$. Закон Лауэ, по сути, это закон релятивистского замедления времени в объеди\-ненном пространстве-времени, которое вовсе не присуще ньютоновской динамике. С точки зрения наблюдателя, подсчитывающего энергетический баланс по законам релятивистской теории, любое инерциальное тело с энергией покоя $E_o=M_o c^2$ теряет при движении часть своей внутренней энергии как при больших скоростях, так и при малых, когда $M_o c^2 {\sqrt {1 - v^2c^{-2}}} -  M_o c^2 \approx   - M_o v^2 /2$. Такой противоток тепловых потерь вдвое снижает величину  кине\-тической энергии поступательного движения термодинамического тела $M_ov^2$  в мире малых скоростей, где $M_ov^2 - M_o v^2 /2 =  M_o v^2 /2 \ll  M_o c^2 $. Еще раз, кинетическая энергия $ M_o v^2 /2$ в модели Ньютона не соответствует энергии медленного поступательного движения $M_ov^2$ в референтах релятивистской физики с замедлением времени.   Так можно ли модельную динамику Ньютона с вдвое заниженной кинетической энергией холодных масс считать правильной теорией \lq\lq{}в области своей применимости\rq\rq{} и подгонять под нее баланс энергий в Специальной (СТО) и Общей (ОТО) теориях относительности? 

\section {Инерция зависит от тепла вопреки Ньютону }

Необходимо признать модельную ущербность и классической механики, и ее понятия массивной точки, для которой нет перспектив конвергенции с носителями энергии в квантовой механике. Ведь эффект замедления времени при переносе  энергии покоя хорошо проверен на примерах  много\-километрового движения пи-мезонов и других короткоживущих частиц. 
От понятия массы покоя $M_o$ надо полностью отказаться в пользу энергии покоя $E_o$ у составного тела или неделимой частицы, поскольку при погло\-щении энергии \lq\lq{}безмассовых\rq\rq{} волн инерция неподвижного тела или атома растет вместе с ростом $E_o$. А сохранение двух понятий  для одной энергети\-ческой характеристики реального термодинамического тела неизбежно приводит сначала к физической путанице, а затем и к неправильным математическим моделям.    Убедимся в этом  на демонстрируемой ниже неспособности ньюто\-новской теории \lq\lq{}холодных\rq\rq{} масс выступать в роли предельного случая для эйнштейновской физики инерциаль\-ных энергий с теплом.

Движение любого термодинамического или составного тела c внутренней энергией покоя $E_o$  описывается  в  метрическом гравитационном поле  \cite {LL} полной релятивистской энергией,
 \begin {equation}
E \equiv \frac {E_o {\sqrt {g_{00}}  }}  {\sqrt {1-\beta^2}} \equiv  {E_o} {\sqrt {{1-\beta^2} } } 
+ \left [ \frac {E_o\beta^2   }  {\sqrt {1-\beta^2}}  - \frac {E_o (1-{\sqrt {g_{00}}  }) }  {\sqrt {1-\beta^2}} \right ], 
\end {equation} 
 которая пропорциональна величине инерциального (и гравитационного) за\-ряда $q \equiv E/\varphi_o$ (где $\varphi_o \equiv c^2/{\sqrt G} = 1.04\times 10^{27}$ эB). В этом  заряде в отсутствие энергии вращения тела мы будем отслеживать  вклады от уменьшающейся со скоростью собственной внутренней энергии термоди\-намической системы $N \equiv {E_o} {\sqrt {{1-\beta^2} }} >0$, от кинетической энергии ее поступательного движения как целого, $K\equiv {E_ov^2   } / {c^2\sqrt {1-\beta^2}} > 0$, и от отрицательной потенциальной энергии  в неоднородном гравитационном поле $ U \equiv $ $- {E_o (1- {\sqrt {g_{00}}  }) }/  {\sqrt {1-\beta^2}} $ $< 0$, где $0 < g_{oo} \leq 1$ и $\beta^2 \equiv v^2/c^2 $   $<1$. Критиковавший Ньютона Эрнст Мах наверняка бы одобрил, что при выбранном определении инерциальных/гра\-витационных зарядов $q$ сила их парного тяготения по закону обратных квадратов зависит  от распределения удаленных тел через  гравитационные вклады в переменный релятивистский заряд $ E/\varphi_o$ каждого из партнеров.

В квадратных скобках в (1) сгруппирована чисто механическая часть энергии поступательно движущегося тела при наличии внешнего гравита\-ционного поля,
 $M \equiv K+U = E-N = E_o (\beta^2 + {\sqrt {g_{00}}} - 1) /  {\sqrt {1-\beta^2}}.$
Кроме того, в разложении по трем степеням свободы полной энергии у термодинамического тела можно объединить кинетические энергии внут\-реннего и внешнего (посту\-пательного) движений, $T \equiv N+K ={E_o } / {\sqrt {1-\beta^2}} $, что и реализовано  СТО в 1905 году  в угоду референтам ньютоновской динамики. Другими словами, СТО в нерелятивистском пределе сориенти\-ровалась на  \lq\lq{}ополови\-ненную\rq\rq{} Ньютоном кинетическую поправку $E_o\beta^2/2$ для поступательного движения и, вопреки своим воз\-можностям, не смогла распознать роль динамических изменений внутрен\-ней энергии при изменений скорости тел и соответствующих замедлений собственного времени.  А ведь скоростные изменения собственной внутренней энергии  ${E_o} {\sqrt {{1-\beta^2} } }$ могли бы наряду с понятием собственного времени выявить  в динамике СТО неразделимый противоток энергии медленного механического движения и сброса тепла, сопутствующего движению в объединенном пространстве-времени. Его метрика  была введена Минковским в том же 1908 году, что и закон  Лауэ. Если не игнорировать инерцию у тепла и закон Лауэ для кинематического уменьшения внутренней энергии, то давно стало бы понятно, что динамика холодных масс Ньютона ни коим образом не должна служить нерелятивистским референтом для физики инерциальных энергий СТО и ОТО, способных объединить  динамику теплых тел с термодинамикой как для общерелятивистского случая, так и для предела малых скоростей в отсутствии полей тяготения. Стал бы ясен и механизм повышенного увели\-чения механической энергии жидкости за счет сброса ее внетренней энергии при начале движения, по новому рассчитывался бы энергообмен в торнадо и других природных явлениях.

\section {Метрическая самодостаточность ОТО с теплыми энергетическими зарядами}
Референты теории Ньютона должны быть отменены не только для динамики  инерционных энергий с теплом в СТО, но и для определения метрических решений через ньютоновский предел для слабого поля ОТО. Действительно, неоднородную метрическую компоненту $g_{oo}(x)$ легко выразить из реляти\-вистского баланса (1) через универсальный гравитационный потенциал $U/E$ для маховского притяжения зарядов полной инерционной энергии $E$, вклю\-чающей тепло: 

\begin{equation}
	{\sqrt {g_{oo}}}\equiv (N+K+U) 
\frac  {\sqrt {1-\beta^2}}  {E_o} \equiv
1 +\frac{U {\sqrt {g_{oo}}}  }{ E} 	\equiv
 \frac {1   }{  [1 +(-U/E)] }.
\end{equation}
	
Основываясь на алгебраических тождествах в соотношениях (1)-(2) можно сформулировать следующую $g_{oo}$-теорему: \lq\lq{}Чисто временная метри\-ческая компонента определяется в гравитации энергий Маха-Эйнштейна непрерывным полевым потенциалом  $\varphi \equiv U/E$ строго как ${g_{oo}} = (1-\varphi)^{-2}$, причем без особенностей на всем полуинтервале  $-\infty < \varphi \leq 0$ допустимых аргументов\rq\rq{}.

Радиальный потенциал $\varphi = -GE_2/c^4 r$  соответствует в слабых полях ньютоновскому тяготению и подчиняется в сильных полях теореме (2) для временной компоненты метрического тензора $g_{oo}(r)= [c^4r/ (c^4r +GE_2)]^2$. Поэтому шварцшильдовская метрика \cite{Sch} с $g_{oo}(r)= (c^2r - 2GM_2)/c^2r$ не вполне согласуется с маховской гравитацией инерциальных энергий (1) с внешними потенциальными вкладами и внутренним теплом.  В энергетике ОТО просто неправомерно равняться на модельную гравитацию холодных масс Ньютона. В 1938 году  Эйнштейн раскритиковал дуальность массивных частиц и безмассовых полей, указав на недуалный путь в физике непрерывных носителей энергии  \cite{EI}. После этого автор ОТО четко охарактеризовал шварц\-шильдовский 4-интервал с сингулярностью как  \lq{}не имеющий отношения к физической реальности\rq{} \cite{Eins} (за что североамериканцами был прозван \lq\lq{}the reluctant father of black holes\rq\rq{}  - отлынивающий папаша черных дыр).

 Напомним, что в 1913 году Эйнштейн и Гроссманн внедрили неоднородное поле во временную часть 4-интервала  пространства-времени Минковского и сохранили, в то время без вариантов, модельную массивную точку или локализованное вещество для референтного сравнения релятивистского и классического движений в гравитационных полях \cite{EG}. После переписки с  Гильбертом у Эйнштейна появилось тензорное Уравнение для гравитационного поля, геометризованного пространственно-временным интервалом $ds= {\sqrt {d\tau^2 - dl^2 }}$ с уже обоими искривленными подъинтервалами, $d\tau^2 \equiv {{g_{oo}}} [dt + (g_{oi} dx^i/g_{oo} )]^2 $ и  
 $dl^2 \equiv (g_{oi}g_{oj}g^{-1/2}_{oo}  - g_{ij})dx^idx^j$. При этом  негеометризо\-ванное вещество в правой части Уравнения Эйнштейна \cite{Ein} опять допускало пространственную локализацию и пустые области в  соответствии с референт\-ными представлениями Ньютона об инерции локализованных масс, а не об инерции непрерывных энергетических полей.      
В чем же состоит нереализованная революционность новых идей Эйнштейна 1938-1939 годов после его первых потрясений основ гравитации в 1913-1916 годах? 

\section {Недуальный поворот от дуальной ОТО }

Метрика Шварцшильда для пустоты не нравилась не только Эйнштейну.  Кривое 3-пространство в метрическом решении 1916 года для постоянного центрального поля \cite{Sch} не могло устроить многих физиков, начиная с отка\-завшегося от дальнейшего соавторства Гроссманна. Зоммерфельд, Швингер, Фейнман и многие другие стремились сохранить евклидовость электродинамики, включая ее  постоянный гауссов поток по двумерной замкнутой поверхности и строгое правило квантования Бора-Зоммерфельда по одномерному замкнутому контуру.  Наиболее далеко в аргументированной критике ОТО 1916 года зашел в 80е годы прошлого века академик А. Логунов,
предложивший в альтернативной теории РТГ \cite{L} вернуться назад к неискри\-вленному прост\-ранству-времени 1908 года и встроить гравитацию в лагранжеву динамику частиц по проверенному сценарию электромагнитных полей. Это бы вернуло релятивистской физике плоское пространство при сохранении референтов ньютоновской динамики        
в \lq\lq{}области совместной применимости теорий\rq\rq{}, устра\-нило бы нефизические черные дыры в сильных полях и привело бы к цикличности состояний материи с высокой и низкой плотностью.

Программные цели Логунова можно только приветствовать, тем более, что Эйнштейн и сам пытался улучшить релятивистскую теорию. При этом он понимал, однако, что последовательная классическая механика, как и квантовая механика протяженных частиц, должна строиться в недуальных терминах плотностей энергии:  \lq\lq{}Вещество  - там, где концентрация энергии велика, поле - там, где концентрация энергии мала. Это различие не качественное, а скорее количественное. Нет смысла рассматривать вещество и поле как два качества, совершенно отличные друг от друга. Мы не можем представить себе определенную поверхность, ясно разделяющую поле и вещество.\rq\rq{} 

Недуальная теория плотностей энергии никак не может трансформироваться в дуальную из-за перехода от высоких концентраций материи к низким.
Неразумно ожидать, что недуальные объекты в физике микромира начнут переходить в дуальные при изменении масштаба пространства или скорости движения частиц. Модельное разделение в макромире единой энергетической сущности непрерывного материального пространства на якобы локализованное вещество и якобы порождаемое им неинерциальное поле объясняется коли\-чественными принципами наблюдений нелокальной реальности, и не может обосновываться как истинное через подстраиваемый под наблюдения мате\-матический аппарат теории (типа неполной динамики Ньютона для холодных массивных точек). 

В 1938 году Эйнштейн предлагал не вернуться от двух понятий 1916 года (пространство-временя-поле плюс вещество) назад  к  сумме трех понятий  (пространство-время плюс поле плюс вещество, как выбрал Логунов \cite{L}), а наоборот сократить число физических понятий до одного, внедрив непрерывно плотность протяженной элементарной частицы в единое пространство-время-поле-вещество. Это революционное предложение не сопровождалось тогда новыми формулами и проверяемыми предсказаниями. Но Эйнштейн сразу предупредил, что в недуальной теории чистого поля непрерывная частица должна исчезнуть из вспомогательной правой части Уравнения 1916 года.  Сегодня каждый может проверить в поддержку недуальных идей Эйнштейна и про-евклидовых целей Логунова, что самосогласованная по энергетическому балансу (1) метрика ОТО с теоремой (2) для $g_{oo}(r) = r^2/(r+r_o)^2$  позволяет по-евклидовски прочесть пространственный подъинтервал $dl^2 = \delta_{ij}dx^idx^j$ при обнулении кривизны Эйнштейна, $G_{oo} = 0$. Это и есть Уравнение Эйнштейна,  $R_{oo} = g_{oo}R/2\neq 0$, без правой части в непустом пространстве статических радиальных плотностей интеграла полной энергии $r_o\varphi_o^2$ для центрального поля тяготения.           

\section {Разворот движения у центра падения}
Для предсказания новых проверяемых эффектов в парадигме непустого пространства Эйнштейна 1938 года надо решительно отбросить в СТО и ОТО ньютоновские референты медленного движения масс в пустоте и перейти по (1)-(2) к подсчету энергетического баланса нагретых тел с учетом переноса тепла.  Достаточно сделать предположение, что релятивистская энергия покоя $E_o$ имеет кинетическую природу за счет внутреннего движения инерциальных плотностей составного тела или частицы. Тогда удаленное от гравитационного центра пробное  тело в покое будет обладать в (1) максимумом свой внутренней кинетической энергии $N(\infty) = E_o$ при нулевой  механической энергии: $M(\infty)=K(\infty)+U(\infty) =0$ при  $K(\infty) = 0, U(\infty)=0$. 
На конечном удалении $R$ от гравитационного центра тело в отсутствие скорости, $\beta(R) = 0$ и $K(R)=0$, продолжает обладать полной внутренней кинетической энергией $N(R) = E_o$  при отрицательной механической энергии $M(R) = U(R) = - E_o [1 - {\sqrt {g_{oo}(R)}}   ] < 0$. При этом полная энергия термодинамического тела, 
$E(R) = N(R) + M(R) =  E_o {\sqrt {g_{oo}(R)}}> 0$, всегда остается в (1) положительной.

Если неподвижное тело освободить на любом расстоянии $R$ от центра гравитации, то по теореме о равнораспределении кинетических энергий по степеням свободы 
оно будет стремится к своему кинетическому равновесию:
\begin {equation}
 N(r_{eq})\equiv {E_o} {\sqrt {{1-\beta^2(r_{eq})} } }  = \frac {E_o\beta^2 (r_{eq})  }  {\sqrt {1-\beta^2(r_{eq})}} \equiv K(r_{eq}),
  \end {equation}
реализующемуся в отсутствии вращений при $v^2(r_{eq}) = c^2/2$ и $r_{eq} < R$.

Без возникновения гравитационных сил не изменялась бы (принцип инерции) и скорость поступательного движения термодинамических тел в сторону кинетического равновесия, т.к. темп изменения кинетической энергии $K$ не может быть в (1) компенсирован лишь темпом потери внутренней энергии $N$.
 Для самопроизвольного набора поступательной скорости и стремления к равновесному распределению внутренних и внешних кинетических энергий инерционные тела взаимным образом создают друг для друга градиенты потенциала и условия для взаимных вращений.
Суть парных сил гравитации в том, чтобы взаимосогласованно обслужить стремление каждого из партнеров к равнораспределению кинетической энергии по всем имеющимся степеням свободы, включая внутренние движения, взаимные поступательные пере\-мещения и встречные вращения.

Вернемся к примеру падения пробной частицы с фиксированной высоты $R$ и нулевой начальной скоростью на покоящийся центр с большой энергией $E_2\gg E_o$, которая создает для полной энергии частицы $E(R)=E_o {\sqrt {g_{oo}(R)}}$ статический гравитационный потенциал через радиальную функцию $g_{oo}(r)$. При свободном радиальном падении в постоянном поле полная энергия частицы (1) по законам ОТО сохраняется, $E[r(t)]/E_o =
 {\sqrt {g_{oo}(r)/[1-\beta^2(r) ]}  } =  {\sqrt {g_{oo}(R)}}$. Такое сохранение позволяет
связать с метрическим полем $g_{oo}(r)$ не только физическую скорость, $0\leq v^2(r) < c^2 $, монотонно набираемую при отвесном падении от нуля до скорости света, но и  координатную скорость тела $dr/dt$ по мировым часам $dt \equiv dx^o/c$ удаленного от центра наблюдателя. Для этого запишем релятивистские соотношения для  строго радиального падения с сохранением энергии пробного тела в постоянном неоднородном поле:
 \begin {equation}
\frac {g_{oo}(r)} {{g_{oo}(R)}} = 1- \frac {v^2(r)}{c^2} = 1 -  \frac {1}{ g_{oo}(r)c^2} \left ( \frac {dr}{ dt} \right )^2.
\end {equation}

С учетом двух направлений движения второе  равенство в (4)  для коор\-динатной скорости можно переписать в эквивалентном виде с введением единичного вектора ${\hat {\bf r}} \equiv {\bf r}/r$:
 \begin {equation}
\frac {d{\bf r}(t)}{dt} =\pm {\hat {\bf r}} c \sqrt {g_{oo}(r)\left [1-  \frac {g_{oo}(r)} {g_{oo}(R)}\right ]}.
\end {equation}
Из этого универсального соотношения ОТО для радиального движения получается, что  с точки зрения наблюдателя тело сначала набирает коор\-динатную скорость, ${\bf 0}\rightarrow d{\bf r}/dt \rightarrow -{\hat {\bf r}}c{\sqrt {g_{oo}(R)}}/2$, в области умеренных полей ${{g_{oo}(R)}}\geq g_{oo} (r\rightarrow r_{eq}) \geq {{g_{oo}(R)}}/2$. Затем, при дальнейшем росте поля по мере движения к центру в области ${{g_{oo}(R)}}/2 \geq g_{oo}(r\rightarrow 0)/ \geq 0$, тело начинает тормозиться до нулевой наблюдаемой скорости,
 $-{\hat {\bf r}}c{\sqrt {g_{oo}(R)}}/2$ $\rightarrow d{\bf r}/dt \rightarrow {\bf 0}$. При этом величина физической скорости $|{\bf v}| = |d{\bf r}/ {\sqrt {g_{oo}(r)}}dt| $ монотонно возрастает от нуля до скорости света на всей траектории сво\-бодного движения к центру поля, причем в конце падения тело полностью теряет собственную энергию покоя $N(0)=0$. Отсутствие энергии покоя и позволяет физической скорости уже безынерционной частицы формально достичь скорости света в самом центре гравитации. На практике всегда измеряется скорость координатного перемещения объекта по мировым часам удаленного от центра наблюдателя. В отличие от физической скорости $v\rightarrow c$, координатная скорость обнуляется в центре, $dr/dt \rightarrow 0$, где наблюдаемое тело временно останавливается перед началом обратного движения.  

После неустойчивой остановки  в крайнем неравновесном состоянии с $N(0)=0, M(0) = K(0) + U(0)=E_o{\sqrt {g_{oo}(R)}}$ в  центре радиального  гравита\-ционного поля  пробное тело начинает обратное ускоренное движение до точки равновесия кинетических энергий $N(r_{eq})=K(r_{eq}) =E_o {\sqrt 2}/2 $  при $v^2 = dr^2/g_{oo}dt^2 = c^2/2$, где вновь достигается максимальная наблюдаемая (координатная) скорость $dr/dt = c{\sqrt {g_{oo}(R)}}/2$ при $g_{oo} (r_{eq})= {{g_{oo}(R)}}/2$. После прохождения по инерции равновесного для тела удаления от центра, $r(t) = r_{eq}$,  тело продолжает набирать внутреннюю энергию $N(r)$ и возвращается с торможением на высоту $R$, где и останавливается в исходном неравновесном состоянии с $E=E_o {\sqrt {g_{oo}(R)}}$, 
$N(R)=E_o, K(R)=0, U(R) / E_o {\sqrt {g_{oo}(R)}}= 1 -  g^{-1/2}_{oo}(R).$

 Чтобы найти наблюдаемое (координатное) ускорение пробного тела на всей траектории свободного падения к центру радиального поля достаточно продифференцировать  координатную скорость (5) по мировому времени 
наблюдателя $t$ и использовать правило сложной производной $dg_{oo}(r[t])/dt $ $=$ $(dg_{oo}/dr) dr/dt $:
\begin {equation}
\frac {d^2 {\bf r}(t)}{dt^2} ={\hat {\bf r}}c^2 \left( \frac {1}{2} - \frac {g_{oo}(r)}{g_{00}(R)} \right)
\frac {dg_{oo}(r) } {dr}   \Rightarrow 
\begin {cases} 
{-(1- \frac {4r_o} {r} )    \frac {c^2 r_o {{\bf r}} }  {r^3}  , \   g^{1916}_{oo} = 1 - \frac{2r_o} {r} \cr\cr
- \frac  {{c^2r_o {\bf r}(r^2-2r_or-r_o^2)}} {(r+r_o)^5},  g^{1938}_{oo} =  \frac{r^2} {(r+r_o)^2}     \cr}
\end {cases}
\end {equation}

При падении на центр с больших удалений, $R/r_o \Rightarrow \infty$,
радиальная зависимость метрики  $g_{oo}(r_{eq}) = 1/2$ вблизи равновесной точки выравнивания кинетических энергий не имеет принципиального значения при описании гравитации теплых тел через теорему о равнораспределении их внутренней и внешней кинетических энергий. Для формального сравнения предсказаний этой теоремы в приложении к тяготению пробных тел мы использовали  в правой части  (6) как общепринятую метрику Шварцшильда 1916 года для пустого пространства вокруг точечного источника, $g_{oo} = 1 - 2r_o/r,$ где $ r_o \equiv GE_2/c^4 $, так и метрическое решение $g_{oo} = r^2/(r+r_o)^2$  из алгебраического тождества (2) с $U/E =-r_o/r$ для недуальной физики энергий 1938 года без пустых для инерции областей пространства. В обоих случаях ньютоновское ускорение слабым полем $ {- c^2 r_o {{\bf r}} }/  {r^3}$ гладко меняет знак на замедление при $r > 2r_o$, т.е. не доходя до предполагаемого шварцшильдовского горизонта в дуальной физике с пустым пространством. 

\section {Выводы по проверяемым предсказаниям}

Общая теория относительности 1916 года с ньютоновскими референтами метрики предсказывает наличие гравитационного горизонта на расстояниях $r = 2r_o\equiv 2GE_2/c^4$ от центра  инерционной энергии $E_2=r_o\varphi^2_o $. Метрические решения ОТО в парадигме непустого пространства инерционных энергий 1938 года предсказывают гладкое уменьшение $g_{oo}(r)$ и наблюдаемое движение пробных тел вплоть до $r=0$. Подтверждение или опровержение наличия горизонта событий в центре галактики может подтвердить или опровергнуть предложение Эйнштейна и Инфельда 1938 года о переходе в ОТО к недуаль\-ному способу описания вещества и поля, что позволило бы реализовать известный с середины прошлого века критерий двойного объединения: частица с полем и электричество  с тяготением \cite{IB}.   

Предсказанный в (6) эффект торможения вертикально падающей материи и ее обратный ускоренный взлет от центра, как бы по вырожденной кеплеровской орбите, мог бы при обнаружении прояснить законы сильного поля и подтвердить природу гравитации через равнораспределение кинетической энергии по степеням свободы. В этом направлении целесообразно провести моделирования наблюдаемого разлета Метагалактики с ускорением на базе предыдущего цикла ее сжатия с торможением.

В рассматриваемом выше свободном падении с остыванием тел по закону Лауэ мы пренебрегали радиационными обменами термодинамического пробного тела с третьими телами или с окружающей материей, играющей роль термостата. На практике повышение внутренней энергии и температуры при уменьшении физической скорости тел (из-за движения тяжелых партнеров или вязких столкновений) сопровождается ростом излучения по закону Стефана-Больцмана. Поэтому объяснение гравитации через теорему о равнораспределении внутренней и внешней кинетических энергий инерциального тела можно проверять посредством наблюдений за космическими объектами. Эта теорема предсказывает, для примера, периодическое выбросы тепла и вулканическую активность нерелятивистски движущихся спутников крупной планеты через их периодические замедления в инерциальной системе удаленных звезд.

Падение на центр галактики из ее периферии характеризуется охлаждением термодинамических тел и их неравновесным поглощением тепла, что подразумевает наличие темной материи для ИК и оптических наблюдений. 
Негеодезическое торможение материи при неупругих столкновениях вблизи центра галактики (равно как и вблизи центра отдельной звезды) приводит к его локальному разогреву  и к запуску термоядерной реакции синтеза легких элементов. 
В звезде энергия термояда за счет теплопроводности  разогревает ее поверхность, что приводит к видимой светимости даже на краю галактики, где остальные падающие тела лишь остывают по закону Лауэ.

Звезды, находящиеся на вытянутых эллиптических орбитах вокруг массивного центра, дважды за цикл тормозятся круговыми сателлитами при минимальных приближениях звезды к этому центру. При этом быстрая потеря поступательной кинетической энергии происходит по закону Лауэ со взрывным  ростом внутренней энергии. Таким образом теорема о равно\-распределении кинетических энергий предсказывает двойной всплеск светимости звезд на каждом витке вытянутой орбиты. При обнаружении такие термокинетические  эффекты могут быть смоделированы для численной проверки изменений внутренней энергии по релятивистскому закону теплопереноса 1908 года. 
 
Релятивистский закон высвобождения внутренней кинетической энергии механических тел или жидкости  важно количественно учитывать даже при малых изменениях их скорости в лаборатории. Это может привести в нерелятивистской гидродинамике и газодинамике к новым термодинамическим вкладам в механическую энергию движущихся сред, что было бы неправильно толковать как нарушение законов сохранения полной энергии динамической системы.

\noindent {\Large \bf {Заряд-энергия с теплом возвращается после отвесного падения на гравитационный центр}}

\bigskip
\noindent{И.Э.Булыженков, bulyzhenkov.ie@mipt.ru} 

\bigskip
\noindent{\it {Физический институт им П.Н. Лебедева РАН\\
Московский физико-технический институт, Москва, Россия} }


\begin{abstract}
Перенос тепла в потоке энергии СТО, а не ток холодной массы Ньютона, дает ОТО правильные референты для геодезики инерциальных энергий и для метрики их непустого пространства. ОТО может сравнить потери внутренней энергии при наборе скорости с ростом энергии поступательного движения  и объяснить тяготение  стремлением тела к равнорас\-пределению своих кинетических энергий по степеням свободы. 
Термо\-динамический подход численно описывает циклическую динамику вертикального падения на центр с торможением на заключительном участке и последущим ускоренным взлетом как осцилляции вокруг кинетического равновесия геодезически дви\-жущегося тела и выдвигает другие проверяемые предсказания для инерциальных зарядов с теплом в гравитации сильных полей.   
\end{abstract}

Ключевые слова: гравитация термодинамических тел, инерция энергии, теорема о равнораспределении, остывание при движении, геодезическое тор\-можение,  
 выброс из центра с ускорением

 \end {document}